\documentstyle[pre,aps,subeqnarray,psfig]{revtex}

\def\Pr{\mbox{Prob}}
\def\var{\mbox{Var}}
\def\ex{\mbox{E}}
\def\cov{\mbox{Cov}}

\title{The Largest Cluster in Subcritical Percolation}

\author{Martin Z. Bazant}
\address{Department of Mathematics, Massachusetts Institute of Technology,
Cambridge, MA}
\date{April 4, 2000}

\begin{document}
\maketitle

\begin{abstract}
The statistical behavior of the size (or mass) of the largest cluster
in subcritical percolation on a finite lattice of size $N$ is
investigated (below the upper critical dimension, presumably $d_c=6$).
It is argued that as $N \rightarrow \infty$ the cumulative
distribution function converges to the Fisher-Tippett (or Gumbel)
distribution $e^{-e^{-z}}$ in a certain weak sense (when suitably
normalized). The mean grows like $s_\xi^* \log N$, where $s_\xi^*(p)$
is a ``crossover size''. The standard deviation is bounded near
$s_\xi^* \pi/\sqrt{6}$ with persistent fluctuations due to
discreteness. These predictions are verified by Monte Carlo
simulations on $d=2$ square lattices of up to $30$ million sites,
which also reveal finite-size scaling.  The results are explained in
terms of a flow in the space of probability distributions as $N
\rightarrow \infty$. The subcritical segment of the physical manifold
($0 < p < p_c$) approaches a line of limit cycles where the flow is
approximately described by a ``renormalization group'' from the
classical theory of extreme order statistics.
\end{abstract}




\section{Introduction}

In the latter half of this century, percolation has become the
canonical model of quenched spatial disorder~\cite{stauffer}. Among
its many areas of application are polymer gelation, hopping conduction
in semiconductors and flow in porous media~\cite{sahimi}.  Percolation
has also attracted the attention of mathematicians because it offers
challenging problems in probability theory of relevance to statistical
physics~\cite{kesten,grimmett}. Since rigorous results are often not
easily obtained, however, computer simulation has played a central
role in the motivation and testing of new theoretical
ideas~\cite{binder}.

Most analytical and numerical studies have examined the critical point
($p = p_c$) where the correlation length $\xi(p)$ diverges, but here
we focus on subcritical percolation ($p < p_c$) characterized by $\xi
< \infty$. In this case, it is known that the cluster-size
distribution $n_s(p)$, the number of clusters of size (or mass) $s$
per site of an infinite hypercubic lattice of coordination $z$, decays
exponentially for all $p< 1/(z-1) < p_c$~\cite{kunz,schwartz}
\begin{equation}
\log n_s \asymp -s \mbox{\ \ as\ \ } s \rightarrow \infty, 
\label{eq:kunz}
\end{equation}
where $a_n \asymp b_n$ means ``$a_n$ scales like $b_n$'', or more
precisely
\begin{equation}
0 < \liminf_{n\rightarrow\infty} \frac{a_n}{b_n} \leq
\limsup_{n\rightarrow\infty} \frac{a_n}{b_n} < \infty.
\end{equation}
(The quantity $P_s = n_s s$, which is the probability that the origin
is part of a cluster of size $s$, is also sometimes called the
``cluster-size distribution'' \cite{kunz,schwartz}.)  The total number
of finite clusters per lattice site at the critical point $n_c \equiv
\sum_{s=1}^\infty n_s(p_c)$ is known analytically for $d=2$ bond
percolation~\cite{temperley,baxter} and numerically for site and bond
percolation for various lattices in $d=2$ and
$d=3$~\cite{lorenz}. Universal finite-size corrections to $n_c$ have
also been studied extensively~\cite{ziff97,kleban,lorenz,ziff99}.

Beyond the rigorous result (\ref{eq:kunz}), it is believed that the
cluster-size distribution decays exponentially for all $p<p_c$ with a
characteristic size $s_\xi(p)$ and a power-law prefactor
\begin{equation}
n_s \asymp s^{-\theta}e^{-s/s_\xi} \mbox{\ \ as\ \ } s/s_\xi
\rightarrow
\infty.
\label{eq:nc}
\end{equation}
where the exponent $\theta$ is supposed to be independent of $p$ with
$\theta=1$ for $d=2$ and $\theta=3/2$ for $d=3$,
respectively~\cite{stauffer}. The quantity $s_\xi$ in (\ref{eq:nc}) is
called the ``crossover size'' since large clusters ($s \gg 1$) of size
much smaller than $s_\xi$ behave ``critically'', while much larger
clusters behave ``subcritically'', as explained below. Because large
clusters are fractal objects, the crossover size and the correlation
length are related by $s_\xi \propto \xi^D$, where $D<d$ is the
fractal dimension of the infinite cluster at $p=p_c$.

In contrast to the cluster-size distribution, relatively little is
known about the size of the {\it largest} cluster $S_{(N)}$ in a {\it
finite} system of size $N = L^d$ for $p<p_c$, with the notable
exception of the recent work of Borgs et al.~\cite{borgs}. (Our
notation for the random variable $S_{(N)}$ is explained below.)  It is
widely believed that the mean largest-cluster size $\mu_N \equiv
\ex[S_{(N)}]$ scales like $\mu_N \propto s_\xi \log N$ for
$p<p_c$. This follows from the heuristic argument $N n_\mu \approx 1$,
which supposes that the largest cluster can be placed independently at
any site in the lattice~\cite{stauffer}. (This useful idea is extended
significantly in section~\ref{sec:simple} below.)  Recently, from
certain scaling axioms verified for $d=2$ and believed to hold for $d
\leq d_c = 6$, Borgs {\it et al.} have proved the somewhat weaker
statement $\mu_{L^d}/\xi^{\prime D} \asymp \log(L/\xi^\prime)$ as
$L/\xi^\prime\rightarrow \infty$, or equivalently
\begin{equation}
\mu_N / s_\xi^\prime \asymp  \log (N/s_\xi^\prime) \mbox{\ \ \ \ as \ }
N/s_\xi \rightarrow \infty
\label{eq:borgs3.1}
\end{equation} 
where $\xi^\prime(p)$ is another correlation length defined in terms
of ``sponge-crossing probabilities'' and $s_\xi^\prime \equiv
\xi^{\prime D}$ is a corresponding crossover size~\cite{borgs}. (Note
that $d\leq d_c$ is assumed throughout this paper.)

In applications $S_{(N)}$ provides a measure of the maximum
connectivity of a random medium, which is of fundamental interest in
the subcritical regime.  From a theoretical point of view, the
``strength'' (or concentration) of the largest cluster $S_{(N)}/N$
plays the role of an order parameter since its expected value in the
``thermodynamic limit''
\begin{equation}
P_\infty(p) = \lim_{N\rightarrow\infty} \mu_N/N
\end{equation}
has a discontinuous slope at $p=p_c$ with $P_\infty(p\leq p_c)=0$ and
$P_\infty(p>p_c)>0$. Beyond the limiting behavior of the mean $\mu_N$,
however, a much more complete understanding of the percolation
transition is contained in the cumulative distribution function
(c.d.f.)  of the largest-cluster size
\begin{equation}
F_N(s) \equiv \Pr(S_{(N)} \leq s)
\end{equation}
which also describes all size-dependent fluctuations of the order
parameter. In this sense, the behavior of $F_N(s)$ near the critical
point fully describes the ``birth of the infinite
cluster''~\cite{borgs}.  Beginning with the same scaling axioms as in
deriving (\ref{eq:borgs3.1}), Borgs et al. have also proved that
$F_N(s)$ varies significantly only on the scale of the mean for
$p<p_c$
\begin{equation}
\lim_{\epsilon\rightarrow 0} \liminf_{N\rightarrow\infty} \left[
F_N\left(\epsilon^{-1}s_\xi^\prime\log(N/s_\xi^\prime)\right) -
F_N\left(\epsilon s_\xi^\prime\log(N/s_\xi^\prime)\right)\right] = 1.
\label{eq:borgs3.5}
\end{equation}
It is believed that (\ref{eq:borgs3.1}) and (\ref{eq:borgs3.5}) would
also hold with the usual definition of $\xi$ as the decay length of
the pair correlation function~\cite{borgs}, so we expect $\xi^\prime
\propto \xi$ and $s_\xi^\prime \propto s_\xi$. 

Although (\ref{eq:borgs3.1}) and (\ref{eq:borgs3.5}) provide important
rigorous justification for the logarithmic scaling of the mean
$\mu_N$, the shape of the distribution $F_N(s)$ and scaling of the
variance $\sigma_N^2 \equiv \var[S_{(N)}]$ appear not to have been
studied (either numerically or analytically) before this
work. Moreover, no connections have yet been made between subcritical
percolation and the classical limit theorems of probability
theory. Such fruitful connections, which are known to explain Gaussian
fluctuations away from the critical point in thermal phase
transitions~\cite{cassandro}, would presumably come from the
statistical theory of
extremes~\cite{frechet,ft,gnedenko,gumbel,galambos}.

The article is organized as follows. First, in order to build the
reader's physical intuition, simple approximations are made in
section~\ref{sec:simple} to derive the asymptotic behavior of $F_N(s)$
and propose finite-size scaling laws for $\mu_N$ and $\sigma_N$. In
section~\ref{sec:numerical}, these predictions are verified for the
$d=2$ square lattice with computer simulations, which also provide
empirical functional forms and numerical parameters for the scaling
laws. Finally, in section~\ref{sec:RG} the preceding results are
explained in terms of a ``subcritical renormalization group''.

\section{Simple Arguments}
\label{sec:simple}

\subsection{Connection with Extreme Order Statistics}

Consider site percolation on a periodic, hypercubic lattice of $N=L^d$
sites. Since any cluster can be uniquely identified with the site
nearest to its center of mass (of lowest index, if there is more than
one such site), we can define a set of $N$ independent, identically
distributed (i.i.d.)  random variables (r.v.) $\{S_i\}$ such that
$S_i=s$ if the largest cluster centered at site $i$ has size $s$ and
$S_i=0$ if no cluster is centered there. Clearly, the most probable
value of $S_i$ is zero, since the number of clusters is always much
less than the number of sites, and it is exceedingly rare to have more
than one cluster centered at the same site, e.g. when one cluster
encircles another.

We seek the c.d.f. $F_N(s)$ of the ``extreme order
statistic''~\cite{gumbel,galambos}
\begin{equation}
S_{(N)} \equiv \max_{1\leq i \leq N} S_i
\label{eq:SN}
\end{equation}
in $N \rightarrow \infty$ with $p<p_c$ fixed. Extreme order statistics
have many classical applications, such as the fracture strength of
solids, the occurrence of manufacturing defects and the frequency of
extreme weather~\cite{gumbel}. More recently in statistical physics,
extreme order statistics have been applied to glassy relaxation on
fractal structures~\cite{rammal}, the dynamics of elastic manifolds in
random media~\cite{vinokur,bm1}, the random energy
model~\cite{bouchaud,saakian}, decaying Burgers
turbulence~\cite{bouchaud}, dispersive transport in amorphous
materials~\cite{kehr} and random sequential
adsorption~\cite{ziffFT}.
In such applications, extreme order statistics are used to describe
the most important features of a random {\it energy} landscape, e.g.
lowest activation energy barriers.

In this work, we show that the theory of extreme order statistics also
has relevance for the {\it geometric} features of random systems. In
one dimension, the largest cluster in percolation bears some
resemblance to the longest increasing subsequence of a random
permutation, which is known to exhibit similar limiting statistics
(see Ref.~\cite{aldous} for a recent review), although the former
problem is much simpler~\cite{1d_note}. Of course, the interesting
cases of percolation, however, are in higher dimensions, which we
address here.

\subsection{A First Approximation Based on Independence} 

The main difficulty in the percolation problem for $S_{(N)}$, aside
from the complexity of the parent distribution, is that the
r.v. $\{S_i\}$ are correlated. Much is known about order statistics of
i.i.d.  r.v.~\cite{gumbel}, but dependent r.v. have been studied
mostly in cases much simpler than percolation~\cite{galambos}.
Nevertheless, considerable insight is gained by neglecting
correlations in deriving an asymptotic form of $F_N(s)$, which will be
justified below in section~\ref{sec:RG}. As one might expect,
correlations in the subcritical regime are too weak to have an effect
in the thermodynamic limit.

Whenever $N \gg s_\xi$, which holds in the limit $p\rightarrow 0$ for
fixed $N \gg 1$, cluster sizes comparable to the system size are
exponentially rare according to (\ref{eq:kunz}).  Since correlations
between the r.v. $\{S_i\}$ arise due to excluded volume effects (see
below), $\cov[S_i,S_j]$ is exponentially small for most pairs of sites
$(i,j)$ in this limit.  Therefore, as a natural first approximation we
assume $N$ independent selections from a continuous parent
distribution with exponential decay
\begin{equation}
\Pr(S_i \leq s) \sim 1 - e^{-s/s_\xi^*} \mbox{\ \ as \ \ } s
\rightarrow \infty
\label{eq:parent}
\end{equation}
where $s_\xi^*(p)$ is an effective crossover size (see below).  Note
that the asymptotic distribution of the maximum of i.i.d. r.v. is
entirely determined by the tail of the parent
distribution~\cite{ft,gnedenko}, so the complicated behavior of $S_i$
for small sizes is irrelevant. From the method of
Cram\'er~\cite{gumbel} applied to (\ref{eq:parent}), we quickly find
\begin{equation}
F_N(s) \sim \left(1 - e^{-s/s_\xi^*}\right)^N
= \left( 1 - \frac{e^{-(s - s_\xi^*\log N)/s_\xi^*}}{N} \right)^N
\label{eq:indep}
\end{equation}
which implies
\begin{equation}
\lim_{N\rightarrow \infty} G_N(z) = e^{-e^{-z}} 
\label{eq:xizero}
\end{equation}
where 
\begin{equation}
G_N(z) \equiv F_N(s_\xi^* z + s_\xi^* \log N) =
\Pr\left(S_{(N)}/s_\xi^* \leq z + \log N\right)
\label{eq:G}
\end{equation}
is a normalized c.d.f. Therefore, in this simple approximation the
largest-cluster size is sampled from the Fisher-Tippett
distribution~\cite{ft_note} with c.d.f.  $e^{-e^{-z}}$, mean $\gamma =
0.5772\ldots$ (Euler's constant) and variance $\pi^2/6$~\cite{ft}; the
mean largest-cluster size grows logarithmically $\mu_N/s_\xi^* \sim
\log N + \gamma$, while the standard deviation converges to a certain
constant $\sigma_N/s_\xi^* \rightarrow \pi/\sqrt{6}$. Comparing with
(\ref{eq:borgs3.1}), we can view the leading-order asymptotic
behavior of the mean 
\begin{equation}
\mu_N \sim s_\xi^*\log N \ \ \mbox{as} \ \ N\rightarrow\infty
\label{eq:mu}
\end{equation}
as defining the effective crossover size $s_\xi^*$ (should it exist),
which is presumably proportional to the others introduced above
$s_\xi^* \propto s_\xi^\prime \propto s_\xi$.

\subsection{Corrections Due to Discreteness} 
\label{sec:discrete}

There appears to be a problem with (\ref{eq:xizero}) for percolation
on a lattice: A discrete c.d.f.  (which is a piecewise constant
function) cannot converge to a continuous function when scaled by a
bounded standard deviation. In fact, since $s$ in
(\ref{eq:parent})--(\ref{eq:indep}) is restricted to integer values,
the limit in (\ref{eq:xizero}) does not exist.  Instead, if we replace
$s$ by $[s]$ (the nearest integer to $s$) in (\ref{eq:parent}), then the
normalized c.d.f. $G_N(z)$ defined by (\ref{eq:G}) approaches a
quasi-periodic sequence of piecewise constant functions with period
roughly $1/s_\xi^*$ in $\log N$,
\begin{equation}
G_N(z) = \left(1 -
\frac{-e^{-z+\delta_N(z)/s_\xi^*}}{N}\right)^N \sim
e^{-e^{-z+\delta_N(z)/s_\xi^*
}}  \ \ \mbox{as} \ \ N\rightarrow\infty
\label{eq:step}
\end{equation}
where 
\begin{equation}
\delta_N(z) \equiv s_\xi^*(z + \log N)  - [s_\xi^*(z + \log
N)].
\end{equation}
(The limiting sequence is strictly periodic only when $e^{1/s_\xi^*}$
is an integer.)  The piecewise constant functions in (\ref{eq:step})
converge weakly in the sense that as $N \rightarrow \infty$ the ``step
edges'' periodically trace out two continuous functions
\begin{subeqnarray}
\overline{G}(z) & \equiv & \limsup_{N\rightarrow \infty}
G_N(z) = e^{-e^{-z-1/(2 s_\xi^*)}} \slabel{eq:limsup} \\
\underline{G}(z) & \equiv & \liminf_{N\rightarrow \infty}
G_N(z) = e^{-e^{-z+1/(2 s_\xi^*)}}
\slabel{eq:liminf}
\label{eq:env}
\end{subeqnarray}
which define a stationary ``envelope'' of width $1/s_\xi^*$ about the
Fisher-Tippett distribution. If we let $a$ be the lattice spacing
(which we take to be unity), then the envelope width would be
$a^d/s_\xi^*$, showing that the lack of convergence is controlled by
the relative importance of discreteness on the scale of the crossover
size.  Note that the continuous distribution (\ref{eq:xizero})
is recovered in the limit $p\rightarrow p_c$ (taken after the limit
$N\rightarrow \infty$)
\begin{equation}
\lim_{p\rightarrow p_c} \overline{G}(z) = 
\lim_{p\rightarrow p_c} \underline{G}(z) = e^{-e^{-z}},
\label{eq:window}
\end{equation}
as the crossover size diverges and hence the envelope width vanishes.

For $s_\xi^*<\infty$ ($p<p_c$), the continuum result for the scaling
of the mean (\ref{eq:mu}) still holds, but the standard deviation has
persistent fluctuations due to discreteness
\begin{equation}
\sigma_N/s_\xi^* \sim \pi/\sqrt{6} + \epsilon_N  \ \ \mbox{as} \ \
N\rightarrow\infty 
\label{eq:sig}
\end{equation}
where $\epsilon_N$ is periodic in $\log N$ with period $1/s_\xi^*$.
Because the limiting sequence (\ref{eq:step}) fluctuates periodically
about a certain fixed distribution, it can be viewed as a ``limit
cycle'' in some appropriate Banach space (see below). Intuitively, the
distribution conforms asymptotically to the Fisher-Tippett
distribution as closely as possible within the constraints imposed by 
discreteness.

\subsection{Corrections due to Correlations} 
\label{sec:corr}

The simple derivation of (\ref{eq:step}) should be valid whenever
$s_\xi \ll 1$ (or $s_\xi^\prime\ll 1$ or $s_\xi^* \ll 1$) because then
even a single site qualifies as a large cluster. If $s_\xi \approx 1$,
however, non-negligible correlations among the r.v. $\{ S_i\}$ arise
because a cluster of size $s_\xi$ excludes on the order of $s_\xi$
nearby sites from being part of any other cluster. If $s_\xi \gg 1$,
on the order of $\xi^d \propto s_\xi^{d/D} \gg s_\xi$ sites are
excluded by such a cluster since it engulfs many smaller, exterior
regions due to its fractal geometry ($D < d$). Therefore, correlations
can be included heuristically by replacing $N$ with $N/s_\xi^\alpha$
in (\ref{eq:step}) which simply shifts the mean by a constant
$\Delta\mu/s_\xi = -\alpha \log s_\xi$ without affecting the
leading-order scaling behavior (\ref{eq:mu}), where $\alpha=0$ if
$s_\xi \ll 1$, $\alpha=1$ if $s_\xi \approx 1$ and $\alpha=d/D$ if
$s_\xi \gg 1$. Note that the effect of correlations is negligible for
$N \gg s_\xi$.  Correlations do, however, control the finite-size
scaling at smaller values of $N$.

\subsection{Finite-Size Scaling} 

There are only two relevant length scales in percolation, the
correlation length $\xi$ and the lattice spacing $a$ (normalized to
unity), or equivalently two mass scales, the crossover size $s_\xi$
(or $s_\xi^\prime$ or $s_\xi^*$) and the volume of a lattice cell
$a^d$ (also normalized to unity). If $s_\xi \gg a$, then discrete
lattice effects on ``large'' clusters with sizes on the order of
$s_\xi$ or larger become negligible, and the system has only one
relevant mass scale $s_\xi$.  As a consequence of the single scale
$s_\xi$ in the limit $p\rightarrow p_c$, any function of $N$ and
$s_\xi$ is expected to collapse into a self-similar form interpolating
between a critical power-law in $N$ valid for $1 \ll N \ll s_\xi$ and
a subcritical function of $N/s_\xi^\alpha$ (for some constant
$\alpha$) valid for $1 \ll s_\xi \ll N$.  For example, because
$\mu_N(p)$ and $\sigma_N(p)$ have the dimensions of $s_\xi$, we have
\begin{subeqnarray}
\mu_N / s_\xi & = & \Phi\left(N/s_\xi^\alpha \right)
\slabel{eq:mu_scale} \\
\sigma_N / s_\xi & = & \Psi\left(N/s_\xi^\alpha \right)
\slabel{eq:sigma_scale}
\label{eq:scale}
\end{subeqnarray}
for some universal functions $\Phi(x)$ and $\Psi(x)$ which do not
depend on $p$. In the critical regime $N \ll s_\xi$, it is expected
that $\mu_N \propto L^D = N^{D/d}$ and that both $\mu_N$ and
$\sigma_N$ are asymptotically independent of $s_\xi$, which implies
$\alpha = d/D$ and $\Phi(x) \propto \Psi(x) \propto s^{D/d}$ as $s
\rightarrow 0$. From (\ref{eq:borgs3.5}) and (\ref{eq:step}), we also
expect $\Phi(x) \asymp \log x$ and $\Psi(x) \asymp 1$ as $x
\rightarrow \infty$.

The classical idea behind the finite-size scaling ansatz
(\ref{eq:scale}) can be understood as follows.  A large subcritical
cluster (on an infinite lattice) intersected with a finite box of side
$L$ exhibits a crossover from ``critical scaling'' at small scales $a
\ll L \ll \xi$ (where a portion of it typically spans the box) to
``subcritical scaling'' at large scales $L \gg \xi$ (where it is
entirely contained within the box). Note that the lattice spacing $a$
is irrelevant as long as $\xi \gg a$; all systems with the same ratio
$L/\xi$ should have equivalent statistics, up to small corrections of
order $a/\xi$ due to discreteness. Of course, as $p \rightarrow 0$ the
finite-size scaling ansatz breaks down, and discrete effects
eventually dominate over correlation effects, as explained above.

\section{Numerical Results} 
\label{sec:numerical}

\subsection{Methods}

In order to test the predictions of the previous section, numerical
simulations are performed for site percolation on periodic $d=2$
square lattices of sizes $N = 5^2$, $13^2$, $31^2$, $74^2$, $129^2$,
$175^2$, $415^2$, $982^2$, $2324^2$ and $5500^2$ with $p = 0.05, 0.1,
0.15, \ldots 0.5$ ~\cite{ap297_note}. Note that the value $p_c =
0.592\ 7460 \pm 0.000\ 0005$ has been determined numerically in this
case~\cite{ziff}.  For each $(N,p)$, between $M=2\times 10^5$ and
$M=10^8$ samples are generated, and clusters are identified by a
recursive ``burning'' algorithm~\cite{binder,burn_note}. With these
methods, trillions of clusters are counted in several months of CPU
time on Silicon Graphics R-10,000 processors.

In performing such large-scale simulations, special attention must be
paid to the choice of (pseudo)random-number
generator~\cite{ziff,ziff2}.  With the standard $32$-bit generator
{\tt rand()}, the largest observed cluster sizes tend to come in
multiples of integers $\geq 2$ (after accumulating data from a very
large number of ``random'' realizations), which indicates that the
periodicity of the generator is having an artificial effect.  In all
the simulations reported here, however, the $48$-bit generator {\tt
drand48()} is used, and the numerical cluster-size distributions
$n_s(p)$ appear to be free of any systematic errors.

\subsection{Largest-Cluster Distributions}

\begin{figure}
\begin{center}
\mbox{ \psfig{file=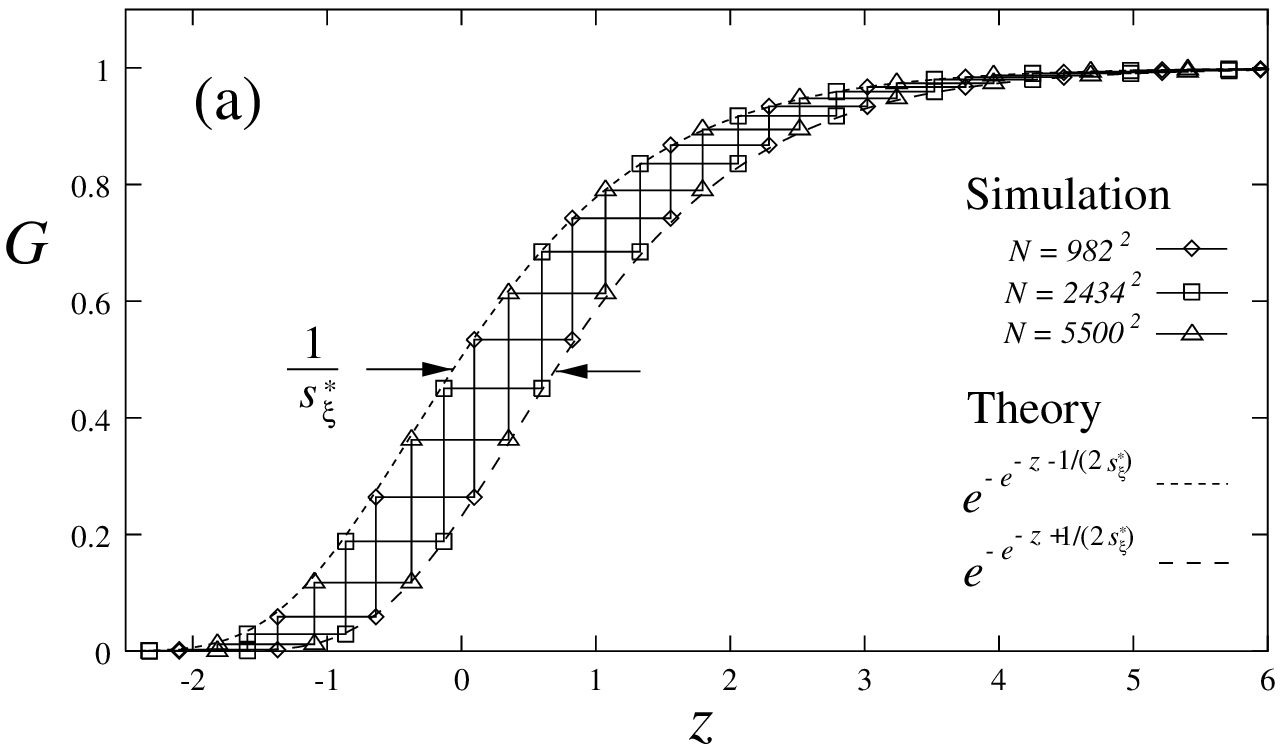,width=5.5in} } \mbox{
\psfig{file=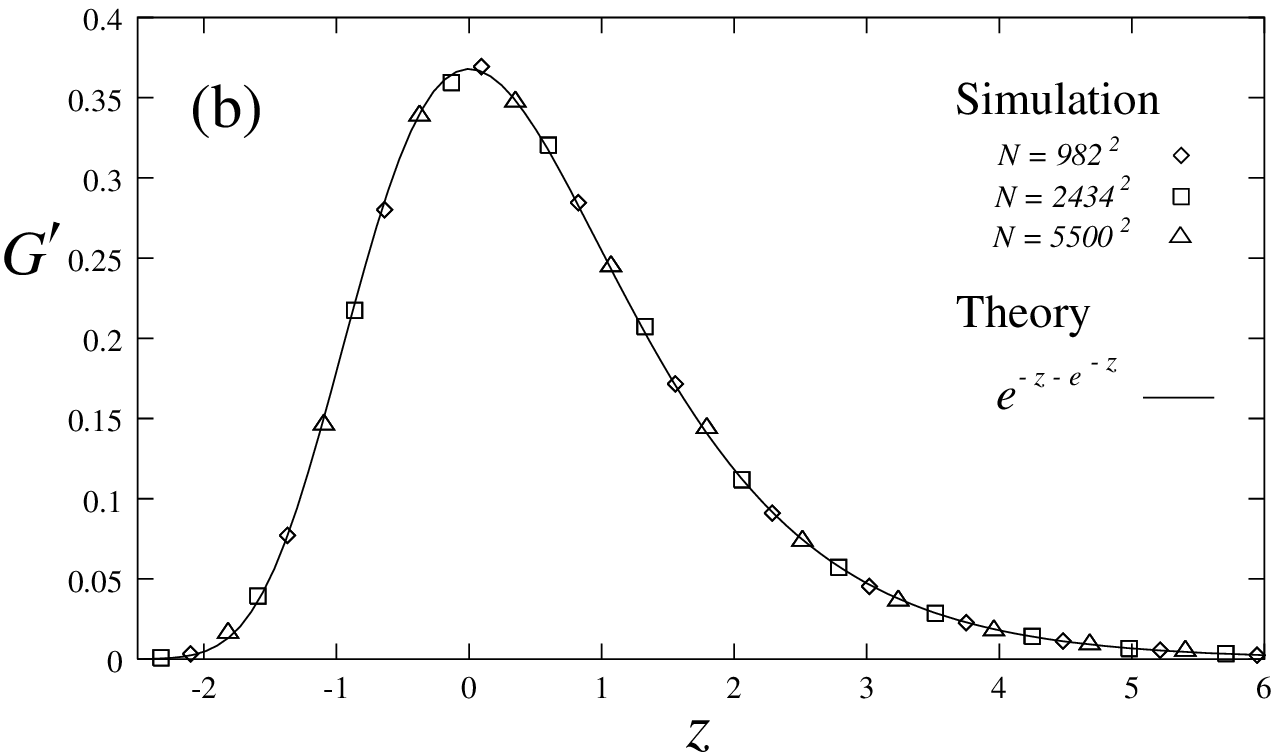,width=5.5in} }
\nopagebreak
\begin{minipage}[h]{5.5in}
\caption{
The discrete c.d.f. in (a) and p.d.f. in (b) of the largest-cluster
size for $p=0.15$ and $N = 982^2$, $2434^2$, $5500^2$, normalized to
have mean $\gamma$ and variance $\pi^2/6$. The c.d.f.s in (a) are
compared with (\protect{\ref{eq:env}}), where
$s_\xi^*=0.90\cdot s_\xi(0.15) = 1.313$.
\label{fig:xvd}
}
\end{minipage}
\end{center}
\end{figure}

The measured largest-cluster distributions are in very close agreement
with the predictions of (\ref{eq:step})--(\ref{eq:env}) for all
$p<p_c$, as shown in Fig.~\ref{fig:xvd} for the case $p=0.15$. In
order to check the shape of the c.d.f. against $e^{-e^{-z}}$, the
distributions are normalized to have mean $\gamma$ and variance
$\pi^2/6$, which differs somewhat from the normalization given above
in (\ref{eq:G}). As predicted by (\ref{eq:env}) the discrete
c.d.f.s in Fig.~\ref{fig:xvd}(a) lie almost perfectly within a
continuous envelope between two Fisher-Tippett
distributions. Likewise, the discrete probability density functions
(p.d.f.) shown in Fig.~\ref{fig:xvd}(b) for $p=0.15$, which are simply
the step heights in Fig.~\ref{fig:xvd}(a), exhibit the expected small
fluctuations about the Fisher-Tippett p.d.f. $e^{-z-e^{-z}}$ due to
discreteness. Using the value $s_\xi^*(0.15) = 1.313$ (determined
independently below), the width of the envelope is seen to be very
close to $1/s_\xi^*$.  Note that the c.d.f.s in Fig.~\ref{fig:xvd}(a)
are shifted slightly outside the envelope by $\epsilon_N\sqrt{6}/(\pi
s_\xi^*)$ because sizes have been scaled by $\sigma_N\sqrt{6}/\pi$
rather than by $s_\xi^*$. Overall, the agreement between
(\ref{eq:step})--(\ref{eq:env}) and the simulation results is
excellent for all the values of $p$ considered here, thus lending some
credence to the approximations of the previous section.

\subsection{Cluster-Size Distributions}

In order to test the finite-size scaling laws (\ref{eq:scale}),
numerical values of the crossover size $s_\xi(p)$ are obtained by
fitting the cluster-size distributions $n_s(p)$ to (\ref{eq:nc}). When
compiling these distributions, unwanted finite-size effects are
minimized by requiring that $N^{1/d}$ exceed the largest observed
cluster size (for a given value of $p$). With this restriction, a
single cluster cannot directly see the periodic boundary
conditions. Motivated by (\ref{eq:nc}), the tails of the cluster-size
distributions are fit to
\begin{equation}
\log n_s = C -\theta \log s - s/s_\xi \mbox{\ \ for\ \ } s > s_{{\small
\mbox{tail}}}.
\end{equation}
Fitting to such an asymptotic form requires some care: The starting
point of the fit $s_{\small \mbox{tail}}$ must be large enough that
the asymptotic behavior is dominant but also small enough that the fit
is not degraded by statistical fluctuations. In this work $s_{\small
\mbox{tail}}$ is systematically chosen where $|ds_\xi/ds_{\small
\mbox{tail}}|$, $|d\theta/ds_{\small \mbox{tail}}|$ and $\chi^2$ are
minimal ($\chi^2\approx 1$). The fit is deemed reliable when the value
thus obtained at $s_{\small \mbox{tail}}$ is contained within all the
other confidence intervals for fits with $s^\prime_{\small
\mbox{tail}}> s_{\small \mbox{tail}}$. Because the raw distributions
have bin counts ranging from over $10^{11}$ at size $1$ down to $0$
and $1$ in the large-size tails, the fitting cannot be done by the
usual least-squares method, which assumes normally distributed
errors. Instead, the parameters $(C,\theta,s_\xi)$ are fit to the
$n_s(p)$ data by Poisson regression, which properly handles the
discrete, rare events in the tail (using the package
X-Lisp-Stat~\cite{xlispstat}).

The fitting results are given in Table~\ref{table:xi}. Fitting errors
grow as $p \rightarrow 0$ because less data is available to accurately
resolve the tail of $n_s(p)$ and also as $p \rightarrow p_c$ due to
critical slowing down.  Although the results for $s_\xi$
should be reliable, the results for $\theta$ (not needed in this work)
could change somewhat if different corrections to scaling were
considered~\cite{binder}. Therefore, the observed small deviation of
$\theta$ from its conjectured value~\cite{stauffer} of $1$ (for all $0
\leq p< p_c$) may only be an artifact of the fit.

\subsection{Scaling of the Mean and Variance}

\begin{figure}
\begin{center}
\mbox{
\psfig{file=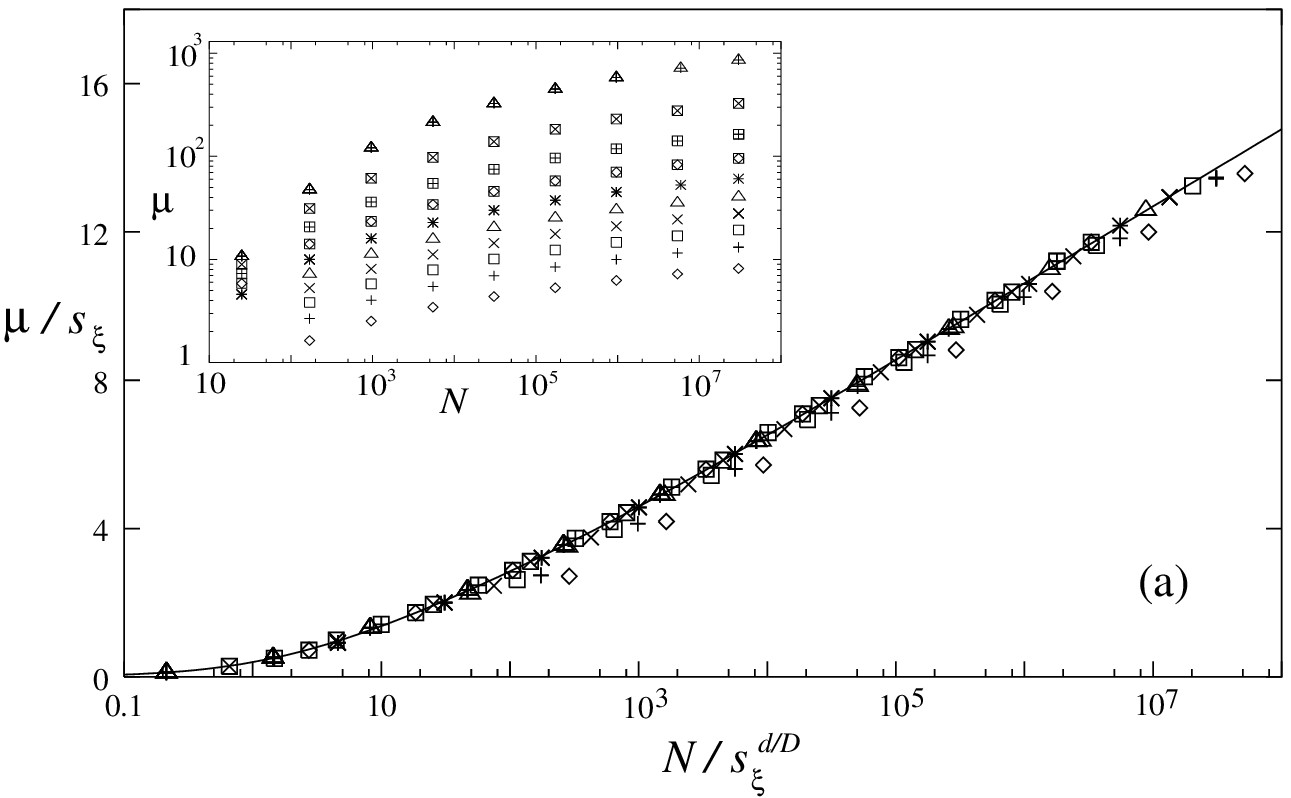,width=5.5in}
}
\mbox{
\psfig{file=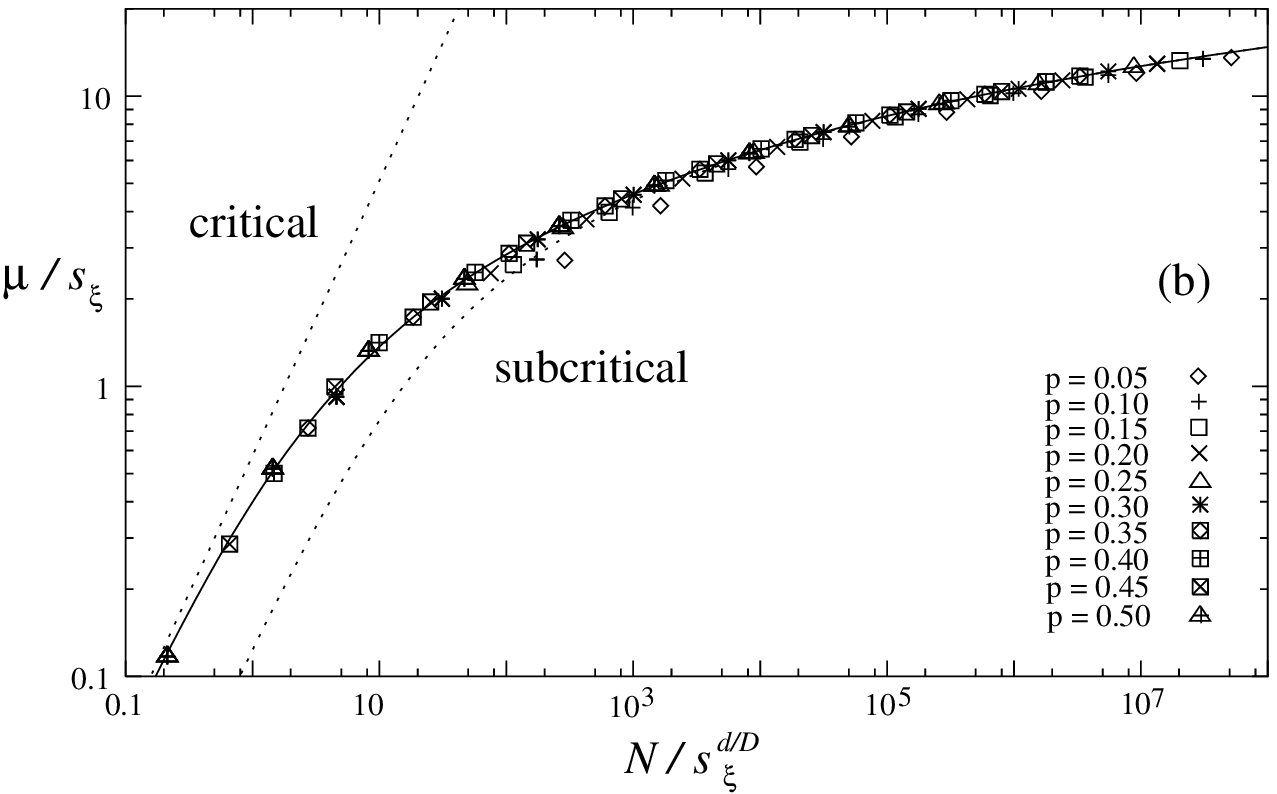,width=5.5in}
}
\begin{minipage}[h]{5.5in}
\nopagebreak
\caption{
The mean largest cluster size plotted as $\mu/s_\xi$ versus
$N/s_\xi^{d/D}$ on a log-linear plot in (a) and a log-log plot in
(b). The solid line fits the $p\geq 0.30$ data to
Eq.~(\protect\ref{eq:mu_fit}) with asymptotic forms given by the
dotted lines. The raw data is in the inset of (a); the legend in (b)
applies throughout.
\label{fig:mu}
}
\end{minipage}
\end{center}
\end{figure}

\begin{figure}
\begin{center}
\mbox{
\psfig{file=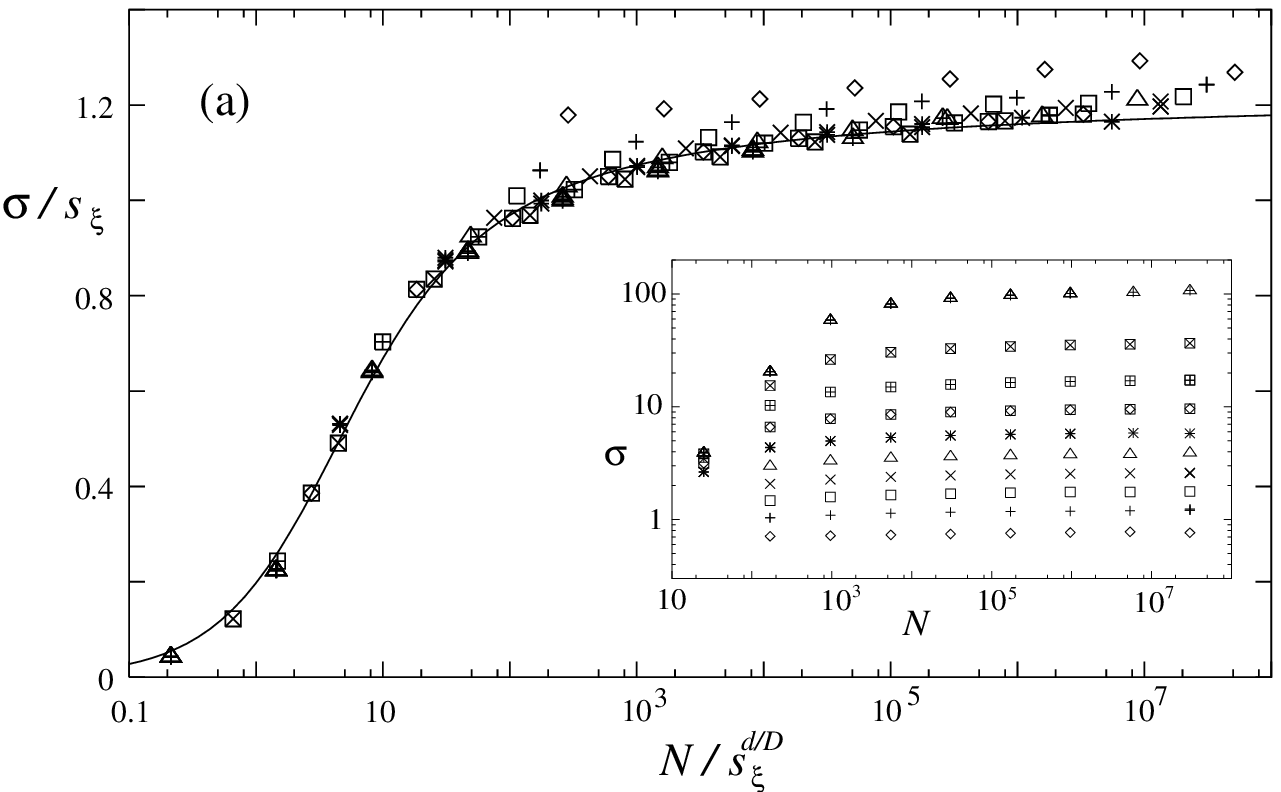,width=5.5in}
}
\mbox{
\psfig{file=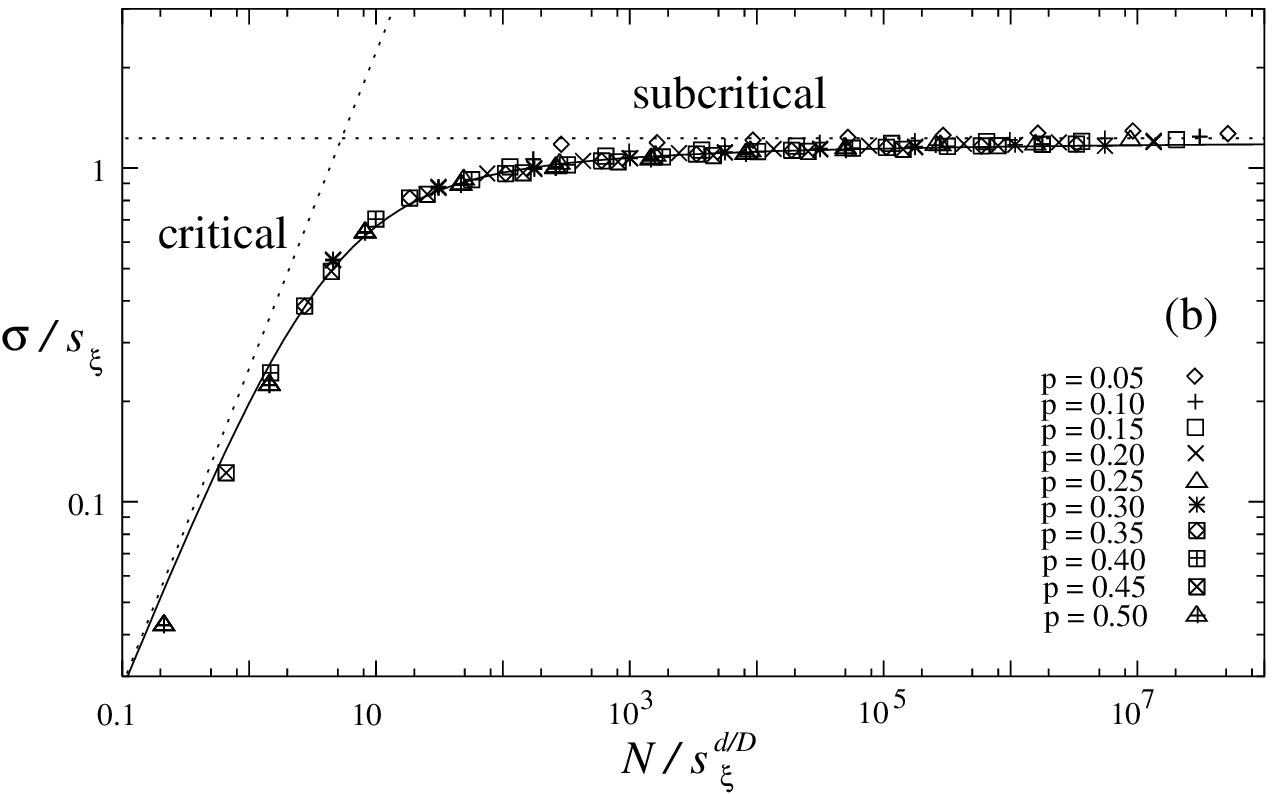,width=5.5in}
}
\nopagebreak
\begin{minipage}[h]{5.5in}
\caption{ The standard deviation of the largest cluster
size plotted exactly as in Fig.~\protect\ref{fig:mu}. In this case,
the $p \geq 0.30$ data is fit to Eq.~(\protect\ref{eq:sigma_fit}).
\label{fig:sigma}
}
\end{minipage}
\end{center}
\end{figure}

\begin{figure}
\begin{center}
\mbox{
\psfig{file=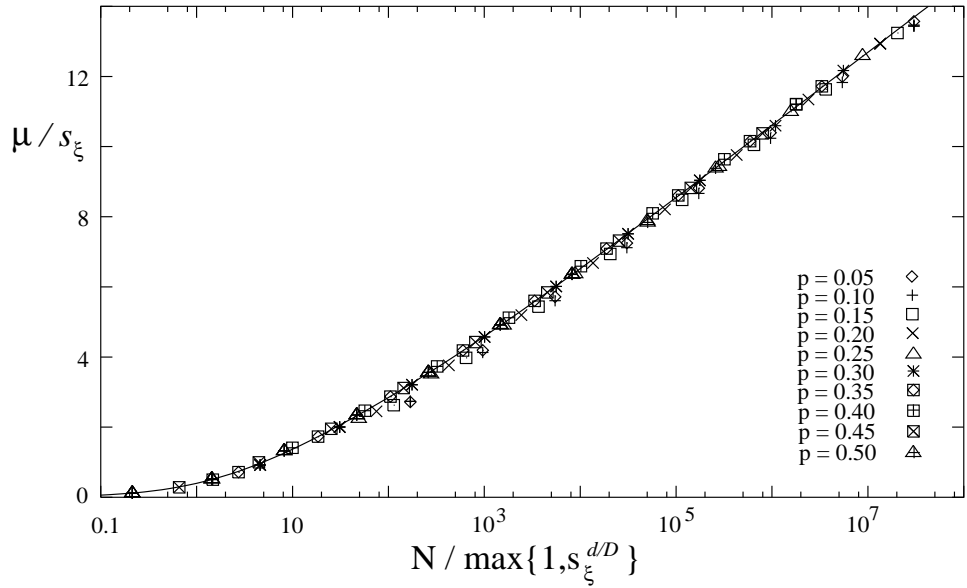,width=5in}
}
\begin{minipage}[h]{5.5in}
\nopagebreak
\caption{
The mean largest-cluster size plotted as $\mu/s_\xi$ versus
$N/\max\{1,s_\xi^{d/D}\}$.
\label{fig:mu_alpha}
}
\end{minipage}
\end{center}
\end{figure}

As shown in Figs.~\ref{fig:mu}--\ref{fig:sigma}, the collapse of the
mean and standard deviation of the largest-cluster size plotted as
$\mu_N/s_\xi$ and $\sigma_N/s_\xi$ versus $N/s_\xi^{d/D}$ (using $D =
91/48$ ~\cite{stauffer}) is nearly perfect for $p \geq 0.30$. As
discrete-lattice effects become important at smaller values of $p$,
however, the data drifts off the universal curves, and the tiny
oscillations predicted by (\ref{eq:sig}) begin to be visible in the
standard deviation.  These effects are most pronounced when $s_\xi <
1$ ($p \leq 0.10$) since then the interpretation of $s_\xi^{d/D}$ as
an excluded volume is meaningless and the second length scale $a=1$
cannot be ignored. Indeed, when $\mu/s_\xi$ and $\sigma/s_\xi$ are
plotted versus $N/\max\{1,s_\xi^{d/D}\}$, as shown in
Fig.~\ref{fig:mu_alpha}, the data for $s_\xi < 1$ lies much closer to
the universal curves, consistent with the heuristic arguments given
above in section~\ref{sec:corr}.

From the simulation results with $s_\xi \gg 1$, the universal scaling
functions in (\ref{eq:scale}) for the $d=2$ square lattice can be
determined numerically.  For $p \geq 0.30$, the scaling function
$\Phi(x)$ for the mean is fit to the empirical form:
\begin{equation}
\Phi(x) = \left[ a_2 + \frac{a_3}{(a_4 + x)^{a_5}} \right]
\log\left[1 + \left(\frac{x}{a_1}\right)^{D/d}\right]
\label{eq:mu_fit}
\end{equation}
where the best parameter values (in the least squares sense) are
$a_1=8.1 \pm 0.5$, $a_2=0.954 \pm 0.005$, $a_3=3.3 \pm 0.2$, $a_4=1.0
\pm 0.3$ and $a_5=0.61 \pm 0.2$.  The collapsed data in
Fig.~\ref{fig:mu} shows a smooth crossover between the expected
critical and subcritical scaling laws, $\Phi(x) \sim 30.3 x^{D/d}$ as
$x \rightarrow 0$ and $\Phi(x) \sim a_2 \log(1 + (x/a_1)^{D/d}) \sim
(a_2 D/d) \log x = 0.90 \log x$ as $x \rightarrow \infty$,
respectively.  The simulation result $\mu_N \sim 0.90 s_\xi
\log N$ justifies our definition of the effective crossover size
$s_\xi^*$ in (\ref{eq:mu}) and for the case of the square lattice
relates it to the crossover size $s_\xi$ in (\ref{eq:nc}) via $s_\xi^*
= 0.90 s_\xi$.

Although the standard deviation appears to be bounded from the data
shown in Fig.~\ref{fig:sigma}, we can only safely conclude $\sigma_N =
o(\log\log N)$ because the subcritical portion of the data only spans
five decades in $N/s_\xi^{d/D}$ (due to memory
restrictions). Following the derivation in the next section, however,
it can be proved~\cite{percsub2} that $\sigma_N = O(1)$ follows from
very reasonable assumptions related to (\ref{eq:borgs3.1}) and
(\ref{eq:borgs3.5}).  Therefore, for $p \geq 0.30$ the scaling
function $\Psi(x)$ for the standard deviation is fit to the empirical
form:
\begin{equation}
\Psi(x) = b_2 \left[ 1 - \frac{1}{1 + b_3 \log\left(1 +
(x/b_1)^{D/d}\right)} \right]
\label{eq:sigma_fit}
\end{equation}
where $b_1 = 8.4 \pm 0.8$, $b_2=1.23 \pm 0.01$ and $b_3=1.5\pm 0.1$.
Once again, as shown in Fig.~\ref{fig:sigma}, the collapsed data for
$s_\xi \gg 1$ fits closely the expected scaling laws $\Psi(x) \sim
0.25 x^{D/d}$ as $x \rightarrow 0$ and $\Psi(x) \sim b_2 = 1.23$ as $x
\rightarrow \infty$. Note that $\sigma/s_\xi^* \sim 1.23/0.90 = 1.36$
for $s_\xi \gg 1$, which differs from $\pi/\sqrt{6} = 1.2825\ldots$ by
only 6.5\%.

\section{Subcritical Renormalization} 
\label{sec:RG}
 
\subsection{Flow in the Space of Distributions} 

There is a profound connection between renormalization-group (RG)
concepts from the theory of critical
phenomena~\cite{fisher,goldenfeld} and the limit theorems of
probability theory through what one might call ``renormalization of
the order parameter'' (as opposed to ``renormalization of the coupling
constant''~\cite{reynolds}). For many second-order phase transitions,
the appropriate order parameter is a sum or average of identical,
correlated r.v. indexed by the sites of a lattice, {\it e.g.}  the
total magnetization in the Ising model.  In such cases the central
limit theorem for i.i.d. r.v.  describes the behavior of the order
parameter away from the critical point, where correlations are
unimportant, and the mathematical concept of a ``stable
distribution''~\cite{levy,feller,bg} amounts to a fixed point of an
RG in the space of probability distributions of the order
parameter~\cite{cassandro,bg}.

In the case of percolation, although the appropriate order parameter
is not the sum but rather the maximum of certain r.v., RG concepts can
still be applied.  Consider the c.d.f. of the largest-cluster size
$F_N(s)$, which we normalize (or rather, successively
``renormalize'') as follows
\begin{equation}
G_N(z) \equiv F_N(\sigma_N z + \mu_N) =
\Pr\left(Z_N \leq z \right),
\label{eq:Gdef}
\end{equation}
where
\begin{equation}
Z_N \equiv \frac{S_{(N)}-\mu_N}{\sigma_N}
\end{equation}
is a r.v. with zero mean and unit variance. Note that since $S_{(N)}$
assumes only integer values, $G_N(z)$ is a piecewise constant function
of $z \in \Re$ with discontinuities at a countable set of points $\{
(s-\mu_N)/\sigma_N | s\in \aleph \}$ with equal spacing $1/\sigma_N$.

\begin{figure}
\begin{center}
\mbox{
\psfig{file=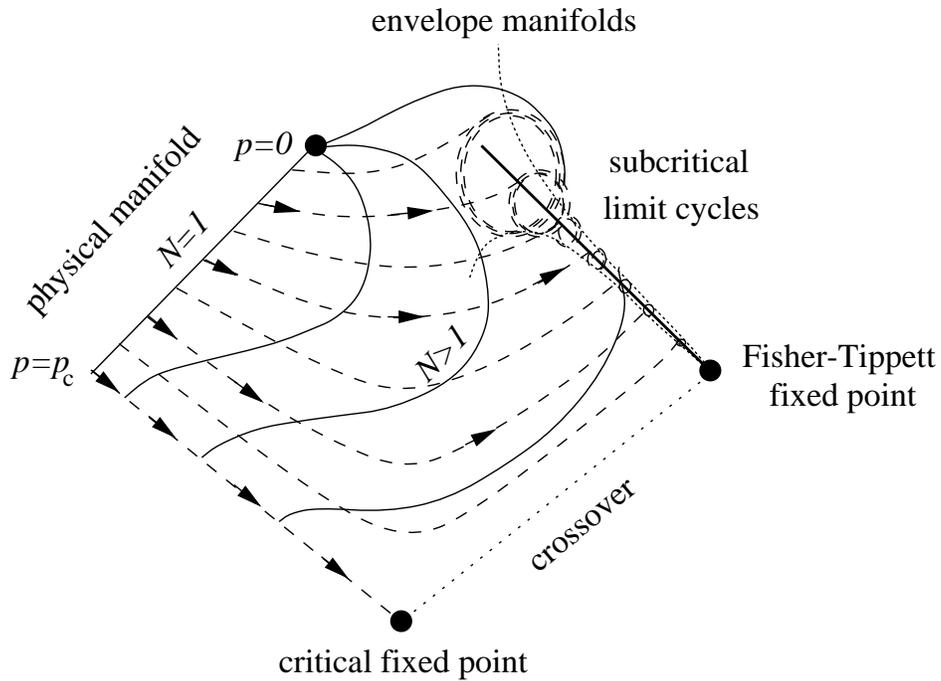,width=5in}
}
\nopagebreak
\vspace{0.2in}
\begin{minipage}[h]{5.5in}
\caption{Sketch of the trajectories (dashed lines) of the normalized
largest-cluster size c.d.f. $G_N(z)$ (in some appropriate function
space) for $p\leq p_c$. Arrows indicate directions of flow as
$N\rightarrow\infty$. The physical manifold is shown for $N=1$ and
three larger values of $N$ (solid lines). Also shown are three fixed
points (thick dots), the subcritical envelope manifolds (short dotted
lines) and the crossover manifold (long dotted line).
\label{fig:RG}
}
\end{minipage}
\end{center}
\end{figure}

The discrete mapping $G_N(z)$ can be viewed as a flow with increasing
$N$ (in some appropriate Banach space, e.g. $L^p$) which advects
distributions towards various possible limiting behaviors. The
subcritical portion of the flow is depicted in Fig.~\ref{fig:RG}. For
each $N\in \aleph$, the set of normalized distributions $\{G_N\}$
parameterized by $0 \leq p \leq 1$ forms a one-dimensional manifold,
which we call the ``physical manifold''.  The ends of the physical
manifold corresponding to $p=0$ and $p=1$ are pinned at trivial fixed
points, which are unit step functions centered at $x=0$ and $x=N$,
respectively (before normalization). Although these fixed points
affect the nearby flow, every trajectory with $0<p<1$ eventually
escapes toward one of three possible limiting behaviors for
sufficiently large $N$: subcritical ($0<p<p_c$), critical ($p=p_c$)
or supercritical ($p_c<p<1$). The latter two cases will be considered
elsewhere; here we focus on subcritical behavior.

According to the heuristic arguments in section~\ref{sec:simple} and
the simulation results in section~\ref{sec:numerical}, the subcritical
segment of the physical manifold is advected into a line of limit
cycles (\ref{eq:G})--(\ref{eq:sig}) around the Fisher-Tippett
distribution once the system size exceeds the crossover size $N \gg
s_\xi$, or more precisely, $N \gg s_\xi^{d/D}$ ($L \gg \xi$).  The
envelope manifolds $\underline G$ and $\overline G$ for $0<p<p_c$
defined in (\ref{eq:env}) enclose the limit cycles. As sketched in
Fig.~\ref{fig:RG}, the ``radius'' of each limit cycle grows as
$p\rightarrow 0$ like $1/s_\xi(p)$, which reflects the influence of
the $p=0$ fixed point representing discreteness. In the opposite limit
$p\rightarrow p_c$
(in the subcritical regime $s_\xi^{d/D} = o(N)$), the envelope
manifolds meet at a fixed point corresponding to the continuous
Fisher-Tippett distribution.

\begin{figure}
\begin{center}
\mbox{
\psfig{file=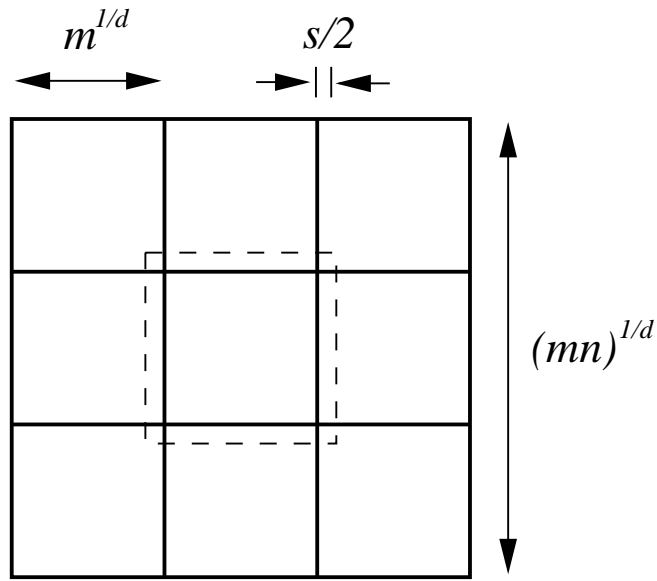,height=3in}
}
\nopagebreak
\vspace{0.2in}
\begin{minipage}[h]{5.5in}
\caption{Sketch of $n=9$ cells (solid lines) used for 
subcritical renormalization on
a square lattice of size $N=mn$.  Along with the nine 
partitioning cells, one enlarged ``supercell''
(dashed lines) used in Ref.~\protect\cite{percsub2} to bound
correlations for cluster sizes smaller than $s$ 
(as described in the main
text) is also shown.
\label{fig:RGpicture}
}
\end{minipage}
\end{center}
\end{figure}

The approach to a fixed point is generally characterized by
self-similarity, which holds ``universally'' for all trajectories
leading to it. In the present case of a lattice-based system, this
asymptotic self-similarity can described by a real-space RG which
relates the c.d.f. $G_N(z)$ for a system of size $N=mn$ to the
c.d.f. for each of $n$ identical, contiguous cells (or blocks) of size
$m$
\begin{equation} 
G_{mn} =  R_{n} G_m
\label{eq:rg}
\end{equation}
in the limit $m\rightarrow\infty$ with $n$ fixed, as shown in
Fig.~\ref{fig:RGpicture}. As usual, the renormalization operators form
an Abelian semigroup under composition $R_{mn} = R_m\circ R_n = R_n
\circ R_m$. These kinds of arguments are typically applied to a  
coupling constant in the vicinity of a critical fixed point, where they
capture the effect of long-range correlations~\cite{goldenfeld}.  They
apply equally well, however, to the order-parameter distribution at a
subcritical fixed point, where correlations disappear.

In a system exhibiting a phase transition, there is a {\it different}
RG of the form (\ref{eq:rg}) valid near each of the various fixed
points. As shown in Figure~\ref{fig:RG}, subcritical trajectories with
$s_\xi \ll 1$ pass by the $p=0$ fixed point and quickly become
ensnared in the subcritical limit cycles, which are described by a RG
given below. Such trajectories never feel much influence from the
critical fixed point because correlation effects are dominated by
discrete-lattice effects, due to proximity of the $p=0$ fixed
point. For larger values of $p<p_c$ such that $1 \ll s_\xi < \infty$,
however, subcritical trajectories at first approach the critical fixed
point ($1 \ll N \ll s_\xi^{d/D}$) before crossing over to the
subcritical limit cycles ($N \gg s_\xi^{d/D}$). This crossover
behavior was demonstrated for the mean and variance above in
Figs.~\ref{fig:mu}--\ref{fig:sigma}, but it also holds for the shape
of the distribution.

In the vicinity of the critical fixed point ($1 \ll N \ll
s_\xi^{d/D}$), trajectories obey a different RG reflecting the
dominance of long-range correlations. The critical fixed point is
unstable in the sense that subcritical and supercritical trajectories
eventually crossover to a different limiting behavior along the
direction of an unstable manifold.
One such ``crossover manifold'' shown in Fig.~\ref{fig:RG}, which
connects the critical and subcritical fixed points, corresponds to the
limits $N\rightarrow\infty$ and $p\rightarrow p_c$ with $N/s_\xi^{d/D}
\rightarrow c$ for some constant $c>0$. Likewise, the stable manifold
converging to the critical fixed point corresponds to the limit
$N\rightarrow\infty$ with $p=p_c$.

\subsection{The Subcritical Renormalization Group}

More than seventy years ago, Fr\'echet~\cite{frechet} and Fisher and
Tippett~\cite{ft} deduced the possible limiting distributions for
extremes for i.i.d. r.v. with the following ingenious argument: If one
partitions $N=mn$ i.i.d. r.v. into $n$ disjoint subsets containing $m$
r.v. each, then the largest of the $mn$ outcomes is equal to the
largest of the $n$ largest outcomes in each subset of size $m$
\begin{equation}
S_{(mn)} = \max_{1\leq i \leq n} S^i_{(m)}
\label{eq:Smn}
\end{equation}
where $S^i_{(m)}$ is the largest outcome in the $i$th subset. Since
the $S_{(m)}^i$ are themselves i.i.d. r.v., the c.d.f. $F_N(s)$ of
$S_{(N)}$ obeys the exact recursion
\begin{equation}
F_{mn}(s) = F_m(s)^n
\label{eq:FRG}
\end{equation}
for all $m$ and $n$ ($m^{1/d}, n^{1/d} \in \aleph$). In terms of the
normalized distribution (\ref{eq:Gdef}), the recursion takes the form
\begin{equation}
G_{mn}(z) = G_m\left(\frac{\sigma_{mn}z + \mu_{mn} -
\mu_m}{\sigma_m}\right)^n, 
\label{eq:GRG}
\end{equation}
which is essentially the subcritical RG for the normalized
largest-cluster size distribution in percolation, but we must also
address correlations and discreteness.  In going from (\ref{eq:FRG})
to (\ref{eq:GRG}) we have defined a ``renormalized'' order-parameter
distribution for percolation valid near the subcritical fixed point,
in much the same way that the Kadanoff-Wilson block-spin construction
defines a renormalized coupling constant for the Ising model valid
near the critical fixed point~\cite{goldenfeld}.

The power of the cell-renormalization approach is that it provides a
natural way to bound correlations and show that the subcritical limit
cycles in percolation are described by the same RG as in the case of
independent random variables (except for the subtle, persistent
fluctuations due to discreteness described earlier). This is
demonstrated rigorously in Ref.~\cite{percsub2}, but here we simply
explain the basic ideas of these authors.  The strategy of the proof
(inspired by Fisher and Tippett) is to fix the number of cells $n>1$
and let the size of each cell $m$ diverge. Since correlations decay
exponentially with distance in the subcritical regime, it seems
plausible that the ``renormalized'' cell random variables $S_{(m)}^i$
would become uncorrelated (as the surface-to-volume ratio vanishes) in
the limit $m\rightarrow\infty$, at least if the dimension were not too
high ($d\leq d_c$).

Precise bounds on the intercell correlations can be obtained as
follows~\cite{percsub2}. If the cells were independent (with free
boundary conditions) we would have $F_{mn}(s) = F_m(s)^n$ as above,
but due to correlations we have instead the upper bound
\begin{equation}
F_{mn}(s) \leq F_m(s)^n
\end{equation}
because joining the $n$ cells together (and thus allowing clusters to
connect and grow) can only increase the size of the largest cluster
(and thus decrease the probability that the largest cluster has size
$\leq s$).  A lower bound can be obtained by considering a set of
``supercells'' (again with free boundary conditions) formed by
appending a ``skin'' of {\it linear} width $s/2$ to each of the
original cells, as shown in Fig.~\ref{fig:RGpicture}. If the mass of
the largest cluster intersected with each of these overlapping
supercells were independently $\leq s$, then the largest cluster
overall would also have mass $\leq s$ (because even a linear chain of
length $s$ would necessarily be completely contained in one
supercell), which yields\cite{FKGnote}
\begin{equation}
F_{(m^{1/d}+s)^d}(s)^n \leq F_{mn}(s) \leq F_m(s)^n.
\label{eq:Fineq}
\end{equation}
These inequalities, which are valid for any dimension $d$, are the
analogs of the Fr\'echet-Fisher-Tippett ``RG'' (\ref{eq:FRG}) for
subcritical percolation, and from them the Fisher-Tippett behavior of
the subcritical limit cycles can be established~\cite{percsub2}.
Heuristically, it is quite plausible that if the ``typical'' largest
cluster size, say within $z$ standard deviations of the mean
\begin{equation}
s_{mn}(z) = \mu_{mn} + z \sigma_{mn},
\end{equation}
does not grow too fast, i.e. $s_{mn}(z) \ll m^{1/d})$ as
$m\rightarrow\infty$ with $n$ and $z$ fixed, then (\ref{eq:Fineq})
should reduce asymptotically to (\ref{eq:FRG}).  Given the results of
Borgs et al. (\ref{eq:borgs3.1})--(\ref{eq:borgs3.5}), we actually
expect the much stronger bound $s_{mn}(z) = O(\log m)$. As explained
below, this logarithmic scaling selects the Fisher-Tippett
distribution $e^{-e^{-z}}$ from among the possible fixed points of
(\ref{eq:FRG}).

\subsection{The Subcritical Fixed Point}

The subcritical fixed point is described by the classical theory for
extremes of i.i.d. r.v.~\cite{gumbel,galambos}. Following Fisher and
Tippett~\cite{ft}, let us assume for now that a continuous fixed point
of (\ref{eq:GRG}) exists pointwise for all $z$, i.e. $G_N(z)
\rightarrow G(z)$ as $N\rightarrow\infty$ and $p\rightarrow p_c$ such
that $s_\xi = o(N)$.  In this case, there must exist finite constants
$a_n>0$ and $b_n$ defined by
\begin{subeqnarray}
\lim_{m\rightarrow \infty}
\frac{\sigma_{mn}}{\sigma_m} =  a_n  \slabel{eq:an} \\
\lim_{m\rightarrow \infty} \frac{\mu_{mn} - \mu_m}{\sigma_m} = b_n 
\slabel{eq:bn}
\label{eq:ab}
\end{subeqnarray}
such that the limiting distribution $G(z)$ obeys the
equation~\cite{ft}
\begin{equation}
G(z) = G(a_n z + b_n)^n
\label{eq:ft}
\end{equation}
which was first discovered by Fisher and Tippett~\cite{frechetnote}.
This functional equation has exactly three solutions, given in
Table~\ref{table:ft}, up to trivial translations and rescalings of $z$
by constants.  In the case of i.i.d. r.v. the basins of attraction of
these three fixed points, which depend only on the tail of the parent
distribution, were first characterized by Gnedenko~\cite{gnedenko}.
In the case of percolation, we have argued above that the appropriate
parent distribution has an exponential tail, which suggests that the
Fisher-Tippett distribution is indeed the subcritical fixed point
(again, ignoring discreteness).

Still assuming that a continuous limiting distribution $G(z)$ exists,
let us make the following additional assumptions
\begin{subeqnarray}
\mu_N & \sim & s_\xi^* \log N \slabel{eq:mscale} \\
\sigma_N - \sigma_{N-1} & = & o(1/N) \slabel{eq:sbound}
\label{eq:scalings}
\end{subeqnarray}
which are clearly supported by our numerical simulations and are
consistent with the rigorous results (\ref{eq:borgs3.1}) and
(\ref{eq:borgs3.5}). These scaling axioms are expected to hold for all
$d\leq d_c$. Note that Eq.~(\ref{eq:sbound}) implies $\sigma_N =
O(\log N) = O(\mu_N)$; with the fact that $\sigma_N$ must be an
increasing sequence, it also implies $a_n=1$ for all $n \in \aleph$
(see \cite{percsub2}).  From (\ref{eq:mscale}) and (\ref{eq:bn}), we
have
\begin{equation}
\frac{\mu_{mn} - \mu_m}{\sigma_m} \sim \frac{s_\xi^* \log n}{\sigma_m}
\rightarrow b_n.
\end{equation}
There are two possibilities: $\sigma_m \rightarrow\infty$ and $\sigma_m
\rightarrow a$ for some constant $a>0$. In the former case, we have
$b_n=0$ and hence $a_n \neq 1$ (see Table~\ref{table:ft}), which is a
contradiction. In the latter case, $b_n = (s_\xi^*/a) \log n$. Without
loss of generality we can set $a=s_\xi^*$ (since this simply amounts
to rescaling $z$) and obtain the equation
\begin{equation}
G(z) = G(z + \log n)^n
\end{equation}
whose only nontrivial solution is $e^{-e^{-z}}$. This also implies that the
standard deviation converges to a constant proportional to the crossover size,
$\sigma_N \rightarrow s_\xi^*\sqrt{\pi}/6$.

\subsection{The Subcritical Limit Cycles}

Of course, the assumption of pointwise convergence to a continuous
limiting distribution is wrong (e.g. see
Fig.~\ref{fig:xvd}). Nevertheless, the conclusions of our simple
derivation are not very different from those of a rigorous analysis
including correlations and discreteness~\cite{percsub2}. Note that
although a limiting distribution $G(z) = \lim G_N(z)$ does not exist,
the envelope functions $\underline{G}(z) = \liminf G_N(z)$ and
$\overline{G}(z) = \limsup G_N(z)$ do exist.  Assuming that
$\underline{G}(z)$ and $\overline{G}(z)$ are continuous (although
$G_N(z)$ is not), it can be shown from (\ref{eq:scalings}) that the
envelope functions have the Fisher-Tippett form
\begin{equation}
\overline{G}(z-z_1) = \underline{G}(z-z_2) = e^{-e^{-z}}
\label{eq:envelope}
\end{equation}
for some constants $-\infty < z_1 \leq z_2 < \infty$ and that the
variance is bounded on the scale of the crossover size
$\sigma_m/s_\xi^* = O(1)$.  The latter result supports our assumption
above in fitting the simulation data to (\ref{eq:sigma_fit}).  The
reader is referred to Ref.~\cite{percsub2} for a detailed proof of
(\ref{eq:envelope}), which follows the RG strategy outlined here. The
heuristic arguments and simulation results in sections
\ref{sec:simple} and \ref{sec:numerical} also lead us to conjecture
that the ``envelope width'' $z_2-z_1$ is simply set by the
``strength'' of the discreteness, i.e. the ratio of the lattice cell
volume ($a^d = 1$) to the crossover size
\begin{equation}
z_2-z_1 = \frac{1}{s_\xi^*},
\label{eq:envwidth}
\end{equation}
which vanishes in the limit $p\rightarrow p_c$.

\section{Conclusion}

In this article, a heuristic theory of the finite-size scaling of the
largest-cluster size in subcritical percolation is presented and
supported by numerical simulations. As expected away from a critical
point, correlations are weak enough that a classical limiting
distribution from the theory of extremes of independent random
variables is recovered once the system size greatly exceeds the
correlation length. This behavior can be easily understood via a
cell-renormalization scheme, which also provides a suitable framework
for rigorous analysis.  Work is underway to extend this work to the
supercritical case, where another classical limiting distribution
arises, and the critical case, which involves a new universality
class.

\section*{Acknowledgments}

The author is grateful to C. Borgs, J. T. Chayes, R. Koteck\'y
and D. B. Wilson for stimulating discussions, E. Kaxiras
and W. C. Carter for computer resources and A. Connor-Sax for help
with X-Lisp-Stat. This work was supported by an NSF infrastructure
grant.



\begin{table}
\begin{center}
\begin{minipage}[h]{5.5in}
\begin{center}
\begin{tabular}{|c|cc|}
$p$ & $s_\xi$ & $\theta$ \\
\hline
\ $0.05$ \ & $0.603 \pm 0.005$ & $1.0 \pm 0.1$  \\
$0.10$ & $0.976 \pm 0.001$ & $0.97 \pm 0.05$ \\
$0.15$ & $1.459 \pm 0.001$ & $0.99 \pm 0.02$ \\
$0.20$ & $2.156 \pm 0.001$ & $1.03 \pm 0.01$ \\
$0.25$ & $3.226 \pm 0.001$ & $1.075 \pm 0.005$ \\
$0.30$ & $4.987 \pm 0.002$ & $1.109 \pm 0.004$ \\
$0.35$ & $8.156 \pm 0.005$ & $1.129 \pm 0.005$ \\
$0.40$ & $14.63 \pm 0.03$  & $1.13 \pm 0.03$ \\
$0.45$ & $31.4 \pm 0.1$    & $1.20 \pm 0.03$ \\
$0.50$ & $91.5 \pm 0.2$    & $1.20 \pm 0.03$ \\
\end{tabular}
\end{center}
\caption{The measured correlation size $s_\xi(p)$ and exponent
$\theta(p)$ for site percolation on the $d=2$ square lattice.
\label{table:xi}
}
\end{minipage}
\end{center}
\end{table}

\begin{table}
\begin{center}
\begin{minipage}[h]{5.5in}
\begin{center}
\begin{tabular}{|cccccc|}
Name & $G(z)$ & Range &\ \ $a_n$ \ \ &\ \ $b_n$\ \ & Basin of Attraction \\
\hline
& & & & & \\
Fr\'echet & $e^{-z^{-\alpha}}$ & $(0,\infty)$
        & $>1$ & 0 & power-law tails \\
& & & & & \\
Weibull & $e^{-(-z)^{\alpha}}$ & $(-\infty,0)$
        & $<1$ & 0 & finite tails \\
& & & & & \\
Fisher-Tippett & $e^{-e^{-z}}$ & $(-\infty,\infty)$
        & $1$ & $\log n$ & exponential tails \\
& & & & & \\
\end{tabular}
\end{center}
\caption{Summary of solutions to the Fisher-Tippett equation. 
In the last column, parent probability distributions $p_1(x)$ in the
basins of attraction of each fixed point are (roughly) described by
their decay at large $x$. (See
Refs.~\protect\cite{gnedenko,gumbel,galambos} for more details.)
\label{table:ft}
}
\end{minipage}
\end{center}
\end{table}


\begin{thebibliography}{99}
\bibitem{1d_note} In one dimension, the largest cluster is simply the
longest sequence of only ``heads'' in $N$ flips of a fair coin.
\bibitem{burn_note} In spite of its suboptimal performance near
the critical point~\cite{stauffer}, the recursive burning algorithm
efficiently identifies the relatively small clusters generated in
subcritical simulations.
\bibitem{frechetnote} Fr\'echet derived the same equation with $b_n=0$
a year earlier~\cite{frechet}. The year before that, L\'evy published
his pioneering work on stable distributions for sums of random
variables~\cite{levy}.
\bibitem{FKGnote} This is technically an application of the FKG
inequality~\cite{grimmett} as explained in Ref.~\cite{percsub2}.
\end{thebibliography}
\end{document}